\documentclass{aastex631}
\usepackage{amsfonts}
\usepackage{amsmath}
\usepackage{amssymb}
\usepackage{graphicx}
\usepackage{color}
\usepackage{natbib}
\usepackage{subfigure}
\usepackage{floatrow}
\usepackage{mathtools}
\usepackage[normalem]{ulem}
\usepackage{soul}

\newcommand{\beq}{\begin{equation}}
\newcommand{\eeq}{\end{equation}}
\newcommand{\beqa}{\begin{eqnarray}}
\newcommand{\eeqa}{\end{eqnarray}}

\shorttitle{Electrostatic Plasma wave excitations at the IP shocks}
\shortauthors{Singh et al.}
\begin{document}
\title{Electrostatic Plasma wave excitations at the interplanetary shocks}
\author[0000-0003-0289-2818]{Manpreet Singh}
\affiliation{Department of Planetary Sciences$-$Lunar and Planetary Laboratory,\\University of Arizona,
Tucson, AZ, 85721, USA.}

\author[0000-0002-5456-4771]{Federico Fraschetti}
\affiliation{Department of Planetary Sciences$-$Lunar and Planetary Laboratory,\\University of Arizona,
Tucson, AZ, 85721, USA.}
\affiliation{Center for Astrophysics$-$Harvard $\&$ Smithsonian,\\
Cambridge, MA, 02138, USA.}

\author[0000-0002-0850-4233]{Joe Giacalone}
\affiliation{Department of Planetary Sciences$-$Lunar and Planetary Laboratory,\\University of Arizona,
Tucson, AZ, 85721, USA.}

\begin{abstract}
 Over the last few decades, different types of plasma waves (e.g., the ion acoustic waves (IAWs), electrostatic solitary waves (ESWs), upper/lower hybrid waves, the Langmuir waves etc.) have been observed in the upstream, downstream and ramp regions of the collisionless interplanetary (IP) shocks. These waves appear as short duration (only a few milliseconds at 1 au) electric field signatures in the in-situ measurements, with typical frequencies $\sim1-10$ kHz. A number of IAW features at the IP shocks seem to be unexplained by kinetic models and requires a new modeling effort. Thus, this paper is dedicated to bridge this gap. In this paper, we model the linear IAWs inside the shock ramp, by devising a novel linearization method of the two-fluid magnetohydrodynamic equations with spatially dependent shock parameters. It is found that, for parallel propagating waves, the linear dispersion relation leads to a finite growth rate dependent on the  the shock density compression ratio, as Wind data suggest. Further analysis reveals that the wave frequency grows towards the downstream
 within the shock ramp, and the wave growth rate  is independent of the electron-to-ion temperature ratio, as Magnetospheric Multiscale (MMS) in-situ measurements suggest, and uniform within the shock ramp. Thus, this study help understand the characteristics of the IAWs at the collisionless IP shocks.

\end{abstract}

\section{Introduction \label{Introduction}}
In collisionless interplanetary (IP) shock waves the upstream plasma parameters evolve to their downstream values on a length scale  much shorter than the interparticle collisional mean free path. In other words, the Coulomb collisions do not influence the properties of the collisionless IP shocks. The IP shocks often show signatures of a wide variety of plasma waves such as the ion acoustic waves (IAWs)
\citep{Fredricks1968,Formisano1982,Balikhin2005,Wilson2007,Goodrich2016,Goodrich2018}, Langmuir waves \citep{Wilson2007}, electrostatic solitary waves (ESWs)
\citep{Williams2005,Wilson2007,Wang2020}, the waves  radiated by the electron cyclotron drift instability (ECDI) \citep{Forslund1971,Forslund1972,Wilson2010}, lower hybrid waves \citep{Walker2008} and the whistler waves \citep{Wilson2016,Wilson2017} in the upstream, ramp and downstream regions. Some waves can show higher occurrence rates than others, e.g., a study by \citet{Wilson2007} suggests that the IAWs are
dominant in the ramp regions of the 67 IP shocks observed in 1997 through 2000 by WIND/WAVES instrument. Thus, this study is focussed on the modeling of the IAWs in the ramp regions of the collisionless IP shock.

Generally, the IAWs are the electrostatic waves which are sustained when the ion mass act as a source of inertia and the electron pressure provides the necessary restoring force. The IAWs can be excited in a magnetized or unmagnetized plasma. In a magnetized plasma, the IAWs can propagate parallel or oblique to the externally applied magnetic field \citep{Yu1980}. It is also found that the angle of propagation of the IAWs with respect to the external magnetic field can influence the properties of these waves \citep{Mishra2013}. These linear IAWs can turn into the Debye-scale electrostatic solitary waves \citep{Wang2020} by nonlinearly trapping the ions.

The observations of the plasma waves associated with the collisionless shocks have a long history. In 1968, \citet{Fredricks1968} reported the first observation of the IAWs in the Earth's bow shock.
In 1979, Gurnett et al. \citep{Gurnett1979} observed the plasma wave turbulence in different regions of the 1976 March 30 IP shock using the Helios 1 and 2 spacecraft data. They observed the three plasma wave modes, characterized as the electron plasma waves, the ion acoustic turbulence and the magnetic noise of whistler mode. It was observed that the ion acoustic activity in the upstream and shock transition regions were more pronounced than that in the downstream region. \citet{Hess1998} analyzed 160 IP shock data encountered by Ulysses spacecraft between 1990 and 1993. In some IP shocks, the IAWs appear hours before and after the shock arrival. In other shocks, IAW activity starts abruptly on the shock arrival. Overall there is no clearly defined pattern for such a wave activity. The probability of wave occurrence (Fig. 2, therein) is highest within a few minutes from the time of shock passage. The occurrence probability of these IAWs were correlated with the ratio of electron-to-proton temperatures ($T_e/T_p$). The authors expected the wave properties to be dependent upon the temperature ratio $T_e/T_p$, which needs to be explained at kinetic scales.  At $T_e/T_p<5$ the damping rate increases significantly with further decrease in $T_e/T_p$ \citep{Fitzpatrick2008}. However, for the data analyzed in \citet{Hess1998}, this ratio always remains $\simeq1$. It was postulated that at small ($T_e/T_p$), the IAWs are excited by the magnetic field aligned proton temperature anisotropies and the electron heat flux.
\par
\citet{Wilson2007} reported a detailed study of the observation of IAWs, ESWs and Langmuir waves via Wind spacecraft in the ramp regions of 67 IP shocks having Mach numbers between 1 and 6. They studied the effect of variation in different parameters on the amplitude of IAWs, and found that the amplitude of the IAWs increases with the increase in shock strength and Mach number. It is also found that the IAWs can generate the anomalous resistivity of the order of 1-856 $\Omega$.m which is important for the dissipation of energy at the IP shocks. \citet{Goodrich2018} reported the oblique Earth bow shock crossing by Magnetospheric Multiscale (MMS) having $\theta_{Bn}\approx43^\circ$ (where $\theta_{Bn}$ is the angle between the normal to the shock surface and the average upstream magnetic field) and Mach number $M_f\approx7$, observed between 06:01-06:05 UTC on 2015 November 27. It was observed that the shock foot have two different regions, one with active magnetic fluctuations and other with no magnetic fluctuations. The region with active magnetic fluctuations showed strong IAWs, ESWs and electron Bernstein wave activity. On the other hand, only the IAWs were prevalent in the quiet region. In a different study \citet{Goodrich2019} analyzed the same shock event as reported in \cite{Goodrich2018} and discussed that the IAWs may be generated due to the impulsively reflected ions. It is also concluded that these waves are of short duration ($10-100$ ms) and short wavelength ($\leq300$ m). \citet{Cohen2020} reported the burst mode STEREO observations of several large amplitude IAWs, ESWs, whistler waves, electron cyclotron drift instability (ECDI) driven waves in the upstream, ramp and downstream regions of 12 quasi-perpendicular IP shocks. It was observed that these waves have highly variable frequency and amplitude. \citet{Davis2021} observed the IAWs, ESWs and the whistler waves in the ramp regions of 11 quasi-perpendicular laminar and perturbed IP shocks. The ramp regions of laminar shocks have higher wave occurrence rates than their downstream regions. The downstream region of perturbed shocks shows 2-3 times more wave occurrence rates than the downstream region of laminar shocks. More recently, \citet{Vech2021} studied the IAWs in the terrestrial foreshock using the MMS data. They employed the interferometric method to find the phase velocity of the IAWs. It was found that the observed group velocity of the waves is less than 40$\%$ smaller than the textbook group velocity \citep{Nicholson1983}. These waves also show time dependent frequency features (on time scales $<100$ ms) possibly due to the presence of reflected ions that lead to IAWs instability.

In the large majority of the observations of IAWs reported so far the electron-to-ion temperature ratio in the upstream region $T_{eu}/T_{iu}$ (where $T_{eu}(T_{iu})$ is the upstream electron (ion) temperature) is found to remain close to 1 or sometimes fall below 1. According to the model in \citet{Nicholson1983,Fitzpatrick2008}, the IAWs are expected to be heavily damped for $T_{eu}/T_{iu}<3$. Thus, it is not well understood how the IAWs are observed despite such a small temperature ratio, when IAWs are expected to damp out quickly before being observed. Thus, this paper aims at addressing this discrepancy.

In the past, several modeling efforts were devoted to relax the constraint of the electron-to-ion temperature ratio threshold by suggesting that in a generic plasma the IAWs can be excited and have a finite growth rate even though $T_e/T_i$ remains $<3$: the instability can arise from temperature gradients in a uniform density plasma as numerically shown by \cite{Allan1974}, velocity shear parallel to the background magnetic field that was theoretically predicted \citep{Gavrishchaka199} and experimentally verified \citep{Agrimson2001}, finite-amplitude whistler wave turbulence that drives intermittently IAWs seen in kinetic scale simulations  \citep{Saito2017}, nonlinear Langmuir wave decay \citep{Zakharov1972a,Zakharov1972b}, counterstreaming ion beam instabilities \citep{Ashour1986}, and the finite electron heat flux \citep{Dum1980}. As for IAWs at shocks, based on a kinetic collisional-regime model \citep{Dum1978a} that includes electrons-IAWs interaction, \citet{Dum1978b} determined the IAWs growth rate as a function of plasma density and electron temperature gradients with an application to perpendicular shocks. In another model for IAWs growth at collisionless shocks, \citet{Priest1972} found that IA instability can be excited by temperature gradient and that the
inhomogeneity of electron density has a small effect: the distortion from Maxwellian of the electron distribution caused by the shock would invalidate the prediction that $T_e/T_i > 3$ is necessary condition for IA instability generation. Moreover, IA instabiity is seen for $T_e/T_i \simeq 1$ also for high Mach number interstellar medium shocks of supernova remnants \citep{Wu.etal:84,Matsumoto2013}. We determine the spatial dependence of the wave frequency and
find a growth rate uniform within the shock ramp.

An analysis of MMS oblique Earth bow shock crossing \citep{Goodrich2018} has shown an upstream deceleration of solar wind ions in a laminar magnetic field region; the presence in this region of reflected ions and electrostatic waves identified as IAWs suggests that the latter mediate the momentum transfer between solar wind ions and reflected ions. At the MMS crossing of the IP shock on 2018 January 8 high-frequency and high-amplitude IAWs have been detected in the ramp and overshoot region likely associated with energized particles  \citep{Cohen2019}. Whether IAWs contribute to the energy dissipation in the plasma foot of laminar shocks, or to the pitch-angle scattering necessary for their early-on acceleration in the shock region \citep{Fraschetti2015,Fraschetti.Giacalone:20}, it is clear that the mechanism originating IAWs at shocks requires a renewed modeling effort.

In this paper we build a new fluid-scale model for the IAWs within the ramp of magnetized shock waves at 1 au and in the near-Sun environment.
For a two-fluid system (electrons and ions) we impose prescribed MHD jump conditions on density compression and velocity of all species but not on the electron-to-proton temperature jump, consistently with most in-situ measurements. The manuscript is structured as follows. In Sec. \ref{Model}, the basic two-fluid model equations are described in context with the IP shocks. In Sec. \ref{dispersion}, the linear dispersion relation of the IAWs is derived within the IP shock ramp. In Sec. \ref{Instability}, the effect of variation in plasma and shock parameters on the characteristics of IAWs and instability is discussed. The Sec. \ref{conclusion} draw the overall conclusion and gives the summary of results.

\section{Model outline \label{Model}}
We consider a perpendicular IP shock moving along the $x-$axis (see Fig.\ref{fig1}) \citep{2021AGUFMSH25G-02S}. The plasma parameters vary smoothly during the shock passage as compared with the duration of the IAWs ($\simeq 10^{-1} - 10$ kHz near 1 au \citep{Wilson2007}).
The magnetic field $\bf B$ is directed along the positive $z-$axis, i.e., ${\bf B}={B_0(x)}\hat{z}$. The plasma is assumed to comprise of positively charged ions (number density $n_i$ and mass $m_i$), and electrons (number density $n_e$ and mass $m_e$). We assume a 2D geometry: the plasma waves  propagate in the $x-z$ plane and spatial variations in the plasma parameters along the $y-$axis are neglected  (i.e., $\frac{\partial}{\partial y}=0$). The basic model is described as follows.

We consider a warm MHD multi-fluid system (ions/electrons) in which a non-zero upstream plasma temperature is assumed ($T_{iu} \neq 0$). Moreover, the shock transition region is assumed to be already formed and stable on the short scale of IAW inverse frequency ($\sim$ ms). A further simplification is that dissipative terms required to maintain the shock structure are neglected. Generally, friction terms can arise from ion reflection at the shock front, electrons trapped behind the shock or the Landau damping. In magnetosonic shocks, a phenomenological friction term proportional to the electrons-to-ions velocity drift along the $y-$axis can be used; from the Ampere equation, such a term is proportional to the magnetic field gradient along the $x-$axis, i.e., average direction of the shock motion. Since at the high IAW frequencies considered here that gradient is typically small, we neglect hereafter any collisionless dissipation term in the momentum and energy conservation equations\footnote{Likewise, anomalous resistivity proportional to $d^2 {v_i}_x / dx^2$ is neglected.}.

The continuity equation, the momentum equation and the equation of internal energy conservation are expressed for multi-fluid non-resistive MHD respectively as:
\beq
\frac{\partial n_j}{\partial t}+\nabla.({n_{\textit{j}}}\bf{{v_\textit{j}}})=0,\label{e1}
\eeq
\beq
\frac{\partial {\bf v_\textit{j}}}{\partial t}+({\bf v_\textit{j}}. \nabla){\bf v_{\textit{j}}}=\frac{Z_\textit{j} e}{m_j}\left(\mathbf{E}+\frac{{\bf v_{\textit{j}}}\times \mathbf{\bf B}}{c}\right)-\frac{1}{m_jn_j}\nabla {P_\textit{j}},\label{e2}
\eeq
and
\beq
\left(\frac{\partial}{\partial t}+({ \bf v_\textit{j}}. \nabla)\right){P_{\textit{j}}}-\gamma_\textit{j}\frac{P_{\textit{j}}}{n_{\textit{j}}}\left(\frac{\partial}{\partial t}+({ \bf v_\textit{j}}. \nabla)\right){n_{\textit{j}}}=0,\label{}
\eeq
where $j=i$ (for ions) and $j=e$ (for electrons), and ${\bf E}$
is the electric field.
\par
For the ion fluid ($j=i$ and $Z_i=+1$) the velocity is given by ${\bf v_\textit{i}}=v_{ix}\hat{x}+v_{iy}\hat{y}+v_{iz}\hat{z}$. For electrons ($j=e$ and $Z_e=-1$), we have ${\bf v_\textit{e}}=v_{ex}\hat{x}+v_{ey}\hat{y}+v_{ez}\hat{z}$.  We also assume that the equation of state for an ideal gas holds for the electrons and ions separately so that $P_{j}=(n_j k_BT_j)$, where $P_{j}$ with $j=i$ or $e$, are the ion or electron pressures, respectively, $T_{j}$ the temperatures and $k_B$ is the Boltzmann constant.

In this paper we focus on the high-frequency limit of IAWs ($1-10$ kHz near $1$ au). In-situ measurements by Solar Orbiter (SolO) of the shock crossing on 2021 November 3 at 0.8 au and a number of interplanetary shock crossings by Wind at 1 au indicate a fluctuating electric field amplitude larger or much larger than the ${\bf v\times B}$ term and the pressure gradient term. As for the former, during the $\sim 1$ minute interval at the SolO shock transition ($\sim 14:04$ UT) the electric field power per unit of frequency $f$, i.e., $P_E = (\delta E^2/8 \pi)/\Delta f$, is of the order of $P_E \sim 10^{-10}$ (V\,m$^{-1}$)$^2$ Hz$^{-1}$ and the magnetic field power per unit of frequency $P_B < 10^{-9}$ (nT)$^2$\,Hz$^{-1}$ (Dr. S. T. Chust and J. Soucek, private communication) that yields $\delta B \lesssim 5 \times 10^{-3}$ nT at 1 kHz. Therefore, the ratio between the fluctuating electric field and the $-(v) \times \delta B$ term (using SI units only for this estimate) can be recast as $\sqrt{P_E/P_B}/v$; for $v\sim 600$ ${\rm km\,s}^{-1}$ (assumed to be equal for protons and electrons for the purpose of this estimate) at $1$ kHz and using $P_B= 10^{-10}$ (nT)$^2$ Hz$^{-1}$, we find $\sqrt{P_E/P_B}/v \sim (10^{-3}\,{\rm V m^{-1}})/ (6\times 10^2 \,{\rm nT \, m\,s^{-1}}) \gg 1$.  Such electric field is larger also than the $-(v/c) \times B_0$, where $B_0 \sim 8$ nT is the unperturbed field strength. The pressure gradient term at SolO ($0.8$ au) for an electron density $n_e\sim8$ cm$^{-3}$ and a downstream electron temperature $\lesssim20\,$eV is of the order of $\sim 10^{-3}$ $\rm V\,m^{-1}$, where we have used a density compression at the shock of $1.5$ (Dr. D. Trotta, private communication) and a temperature jump across the shock equal to $5$; thus, the pressure gradient term is smaller than the fluctuating electric field amplitude at high frequencies.

Since the measured high-frequency electric field is larger than the motional electric field and pressure gradient contribution, such an electric field must originate from the short-lived charge imbalance between electrons and ions: the two-fluid MHD equations are to be coupled with the Poisson equation
\beq
\nabla . {\bf E}=4\pi \rho=4\pi e(n_i-n_e),  \label{poisson1}
\eeq
where $\rho=e(n_i-n_e)$ is the net local charge density in the plasma.

In the electrostatic approximations \citep{Stix1992}, the non-zero electric field components can be written as:
\beq
E_x=-\frac{\partial \phi}{\partial x}, \quad \mathrm{and}  \quad E_z=-\frac{\partial \phi}{\partial z},
\eeq
where $\phi$ is the electrostatic potential. Thus, taking into consideration all the above assumptions, we can write the ion and electron fluid equations together with the Poisson equation in the component form. For ions:
\beq
\frac{\partial n_i}{\partial t}+\frac{\partial (n_i v_{ix})}{\partial x}+\frac{\partial (n_i v_{iz})}{\partial z}=0,\label{ioncontinuity}
\eeq
\beq
\frac{\partial v_{ix}}{\partial t}+v_{ix}\frac{\partial v_{ix}}{\partial x}+v_{iz}\frac{\partial v_{ix}}{\partial z}=-\frac{e}{m_i}\frac{\partial \phi}{\partial x}-\frac{1}{m_i n_i}\frac{\partial (n_i k_B T_{i})}{\partial x}+\Omega_i v_{iy},\label{ionmomentx}
\eeq

\beq
\frac{\partial v_{iy}}{\partial t}+v_{ix}\frac{\partial v_{iy}}{\partial x}+v_{iz}\frac{\partial v_{iy}}{\partial z}=-\Omega_i v_{ix},\label{ionmomenty}
\eeq

\beq
\frac{\partial v_{iz}}{\partial t}+v_{ix}\frac{\partial v_{iz}}{\partial x}+v_{iz}\frac{\partial v_{iz}}{\partial z}=-\frac{e}{m_i}\frac{\partial \phi}{\partial z},\label{ionmomentz}
\eeq
where $\Omega_i=\Omega_i(x)={e B_0(x)}/{m_i c}$ is the ion cyclotron frequency.

For electrons, on neglecting the electron inertia (i.e., $m_e\rightarrow0$),
the $x$ component of electron momentum equation could yield $E_x$, as a generalized Ohm equation (the finite electron mass is only important for $E_y$, $E_z$, since the gradient of ${v_e}_z$ can be large). The two-fluid approach in Eq. (\ref{e2}) leads to
\beq
0=E_x+\frac{B_0}{c} v_{ey}+\frac{1}{ en_e}\frac{\partial (n_e k_B T_{e})}{\partial x},\label{electronmomentx}\nonumber
\eeq
\beq
0=E_y+\frac{B_0}{c} v_{ex},\label{elemomenty}\nonumber
\eeq
\beq
0=-E_z-\frac{1}{en_e}\frac{\partial (n_e k_B T_{e})}{\partial z}.\label{electronmomentEz}
\eeq
However, as shown above, in the high-frequency regime, i.e. $\sim 1\,$kHz, the fluctuating electric field measured by SolO is larger than the $v\times B$ term.
{Thus, the measured high-frequency electric field cannot be the one provided by the generalized Ohm's law (using the electron momentum equation), but the electrostatic regime (using the Poisson equation) determines $E$.

The Eq. (\ref{electronmomentPhi}) can be written in a more useful form:
\beq
0=-\frac{\partial \phi}{\partial z}+\frac{1}{en_e}\frac{\partial (n_e k_B T_{e})}{\partial z}.\label{electronmomentPhi}
\eeq
The electron spatial (time) scales are very small (fast) as compared to the ion spatial (time) scales. Since we are looking at the wave formation at the ion spatial (time) scales, the electrons can just be assumed as a thermal background. This assumption is based on the fact that the electrons have a relatively small mass as compared to that of the ions; thus, the electrons can oscillate faster, generate a thermal background, and provide the restoring force for the formation of IAWs.
Thus, we can keep terms of order $v\times B$ in the ion momentum  equations.

In such a warm plasma the conservation of internal energy needs to be included. We neglect here the heat flux guided by the following measurements: MMS data of the Earth's bow shock crossing of 2019 March 05 show both upstream and downstream very large fluctuations of the electron heat flux and a value along the shock normal consistent with a negligible heat flux within the shock ramp with respect to the kinetic, magnetic and enthalpy fluxes \citep[Fig.3 in][]{Schwartz2022}. We also neglect heat exchange between ions and electrons. The anisotropy of the pressure tensor is also neglected:  small components of the magnetic field normal to the shock lead to anisotropies in the momentum particle distribution that might lead to the mirror mode instabilities
likely observed at the Voyager 1 crossing of the termination shock \citep{Liu2007}. Wind measurements \citep{Vogl2001} at the Earth's bow shock crossing show a small variation in plasma parameters (e.g., density) that can be accounted for by pressure anisotropies. However, for nearly supercritical (Mach numbers $\lesssim 5$) perpendicular shocks, the change in the plasma parameters is relatively small.
The internal energy conservation for ions along the direction of motion of the shock, i.e., along the $x-$axis, is written as
\beq
\frac{\partial (n_i k_BT_i)}{\partial t}+{v_{ix}}\frac{\partial (n_i k_BT_i)}{\partial x}-\gamma_\textit{i}k_BT_i\left(\frac{\partial n_i}{\partial t}+{v_{ix}}\frac{\partial n_i}{\partial x}\right)=0.\label{eq1a}
\eeq
The electron energy conservation is written as
\beq
\frac{\partial (n_e k_BT_e)}{\partial t}-\gamma_\textit{e}k_BT_e\left(\frac{\partial n_e}{\partial t}\right)=0.\label{ele-energy}
\eeq
It is important to notice that the Eq. (\ref{ele-energy}) does not contain the spatial pressure gradients: for the electron velocity component $v_{ex}\, \partial P_e/\partial x \sim v_x P_e/(c/\omega_{pi}) \sim (10^7 {\rm cm \,s^{-1}})/10^7$ cm $\sim 1$ s$^{-1} \times P_e$ where the length scale of pressure gradient is taken to be the ion inertial length $c/\omega_{pi} \sim 10^7$ cm at $1$ au ($c$ is the velocity of light and $\omega_{pi}=\sqrt{\frac{4\pi n_{iu} e^2}{m_i}}$ is the upstream ion plasma frequency); this term is much smaller than the time variation term $\partial P_e/\partial t \sim P_e/\omega \sim 1$ m\,s$^{-1} \times P_e$. Also, the gradients of $n_e$ and $k_BT_e$ along the $y-$ and $z-$axis are vanishing, hence eq. \ref{ele-energy}.

The most important equation under these assumptions is the Poisson equation which closes system, rewritten as:
\beq
(n_e-n_i)=\frac{1}{4\pi e }\left(\frac{\partial ^2 \phi}{\partial x^2}+\frac{\partial ^2 \phi}{\partial z^2}\right).\label{poisson}
\eeq
From the above equation, it can be seen that
the instantaneous differences {(at a given location)} in fluctuations in the ion and electron number densities generate the fluctuating electric field. In the following section we perform a linear perturbation analysis of the equations listed above.

\section{Linear analysis\label{dispersion}}
\subsection{Shock parameters variation across the ramp}
The linearization scheme is illustrated in Fig. \ref{fig1}. It can be seen that the background plasma parameters (e.g., the ion number density $n_{i0}$) upstream of the shock are assumed to be uniform, despite the routinely measured density inhomogeneity: from in-situ measurements, the upstream self-generated turbulence, if identified, is expected to drown far from the shock into the pre-existing turbulence \citep[e.g.,][]{Burlaga.etal:06b}. The effect of the two combined turbulence sources on the momentum spectrum of the energetic particles at the shock has been modeled in \cite{Fraschetti2021,Do.etal:21,Fraschetti.Balkanski:22}. Similarly, we can model the other background plasma parameters such as the electron number density $n_{e0}$, ion velocity along $x-$axis ($v_{ix0}$) and the ion(electron) temperature $T_{i0}(T_{e0})$. The shock is assumed to be infinitely planar: for all zero-th order plasma observables (density, temperature, ion velocity along the $x-$axis) it holds $\partial(\cdot)/\partial z =0$. Thus, the values of these parameters in different regions, and the linearization scheme can be expressed as given below

\DeclarePairedDelimiter\Floor\lfloor\rfloor
\DeclarePairedDelimiter\Ceil\lceil\rceil
\[
  \Re(x) =
  \begin{cases}
  \Re_0'                                   & \text{for $x<x_0$}\hspace{0.16cm}\quad \quad \quad \mathrm{(Far \quad upstream),}\\
  {\Re_0''}                                  & \text{for $x_0<x<x_1$} \quad \mathrm{(Foot \quad region),}\\
  \Re_0''+\sigma x+\Re_1                 &\text{for $x_1<x<x_2$}\quad \mathrm{(Ramp \quad region),}\\
  \Re_d                                    &\text{for $x>x_2$}\hspace{0.16cm}\quad\quad\quad \mathrm{(Downstream),}
  \end{cases}
\]
where $\Re=n_{i}, n_{e}, v_{ix}, T_{e}, T_{i}$, and $\sigma$ is the slope corresponding to each of these parameters within the ramp. The term $\Re_1$ represents the small amplitude linear fluctuations in plasma parameters. The very small variation of the background plasma parameters in the foot region can be neglected here. So, practically, we can consider $\Re_0'\approx\Re_0''\equiv\Re_u$ (e.g., $n_{i0}'\approx n_{i0}''\equiv n_{iu}$),  where $\Re_{u}$ is the value just ahead of the shock ramp.

\begin{figure}[hb]
\centering
\includegraphics[width=0.75\textwidth]{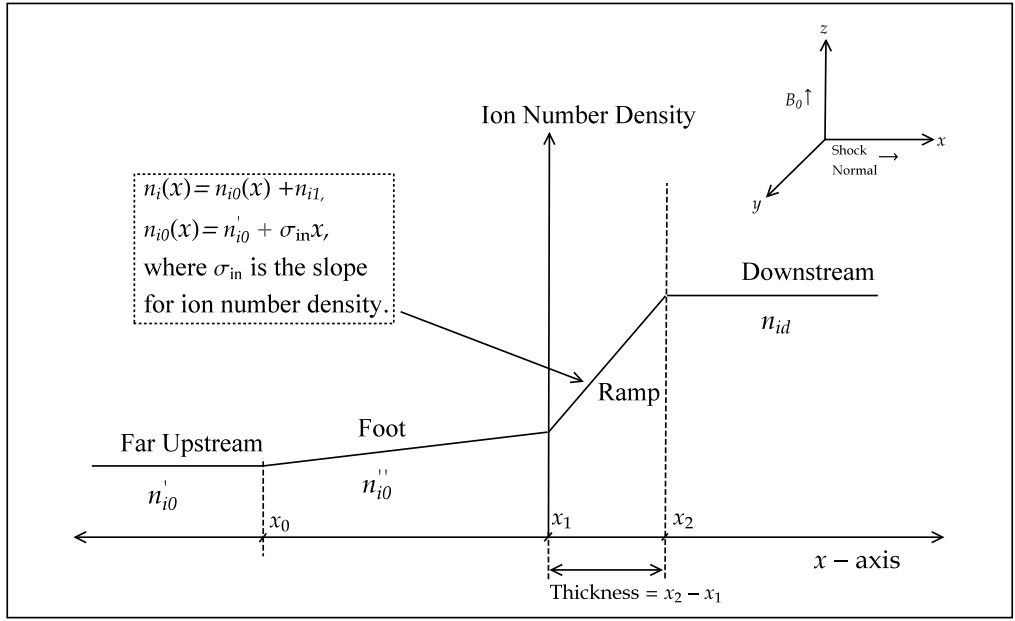}
{\caption{{A simple illustration of the profile of the ion number density across the shock.}}\label{fig1}}
\end{figure}
The slope within the ramp ($\sigma_{in}$) for $n_{i0}(x)$ can be defined as
\beq
\sigma_{in}=\frac{(n_{id}-n_{iu})}{(x_2-x_1)}
=n_{iu}\frac{(r_{in}-1)}{(x_2-x_1)} ,
\eeq
where $r_{in}=\frac{n_{id}}{n_{iu}}$ is the jump in the ion number density. Here $(x_2-x_1)\approx\frac{c}{\omega_{pi}}$ represents the scale for the thickness of the shock ramp.
Observations show that although typically the shock ramp thickness is of the order of a few electron inertial lengths (i.e, $\sim5-15$ $c/w_{pe}$),  occasionally the ramp thickness can reach ion inertial length \citep{Newbury1998,Hobara2010,Mazelle2010}. Assuming a shock ramp thickness of the order of $\sim5-15\,c/\omega_{pe}$, the wavelength of IAWs will still be significantly smaller than the ramp thickness.
Similarly, for $n_{e0}(x)$, $v_{ix0}(x)$, $T_{i0}(x)$ and $T_{e0}(x)$, we can respectively write the gradients as
\beq
\sigma_{en}=n_{eu}(r_{en}-1)\frac{\omega_{pi}}{c} \quad \mathrm{where}\quad r_{en}=\frac{n_{ed}}{n_{eu}},
\nonumber
\eeq
\beq
\sigma_{iv}=v^{sh}_{ixu}(r_{iv}-1)\frac{\omega_{pi}}{c} \quad \mathrm{where}\quad r_{iv}=\frac{v^{sh}_{ixd}}{v^{sh}_{ixu}}=\frac{1}{r_{in}},\label{sig-riv}
\eeq
\beq
\sigma_{it}=k_B T_{iu}(r_{it}-1)\frac{\omega_{pi}}{c} \quad \mathrm{where}\quad r_{it}=\frac{T_{id}}{T_{iu}},
\nonumber
\eeq
\beq
\sigma_{et}=k_B T_{eu}(r_{et}-1)\frac{\omega_{pi}}{c} \quad \mathrm{where}\quad r_{et}=\frac{T_{ed}}{T_{eu}} \, ,
\nonumber
\eeq
where
$v^{sh}_{ixu}$ and $v^{sh}_{ixd}$ are the $x-$components of the ion velocity in the shock frame, respectively in the upstream and downstream regions, and in the definition of $\sigma_{iv}$ we have used the mass conservation from the jump conditions.
Following the above schematic, the complete set of linearized parameters can be written as
\beq
n_i(x,z)=n_{i0}(x)+n_{i1}(x,z), \quad n_e(x,z)=n_{e0}(x)+n_{e1}(x,z),
\eeq
\beq
\quad v_{ix}(x,z)=v_{ix0}(x)+v_{ix1}(x,z), \quad v_{iy}(x,z)=v_{iy0}+v_{iy1}(x,z), \quad v_{iz}(x,z)=v_{iz0}+v_{iz1}(x,z)
\eeq
\beq
v_{ez}(x,z)=v_{ez0}+v_{ez1}(x,z), \quad {\phi(x,z)=\phi_0(x)+\phi_1 (x,z)}
\eeq
\beq
k_BT_i(x,z)=k_BT_{i0}(x)+k_BT_{i1}(x,z), \quad k_BT_e(x,z)=k_BT_{e0}(x)+k_BT_{e1}(x,z)
\eeq
where the zero-th order components are decomposed into the uniform part and the part spatially varying within the ramp:
\beq
n_{i0}(x)=n_{iu}+\sigma_{in} x, \quad n_{e0}(x)=n_{eu}+\sigma_{en} x,
\eeq
\beq
\quad v_{ix0}(x)=v_{ixu}+\sigma_{iv}x, \quad v_{iy0}=v_{iyu}, \quad v_{iz0}=v_{izu},
\eeq
\beq
v_{ez0}=v_{ezu}, \quad \phi_0(x)=\phi_u,
\eeq
and
\beq
k_BT_{i0}(x)=k_BT_{iu}+\sigma_{it}x, \quad k_BT_{e0}(x)=k_BT_{eu}+\sigma_{et}x.
\eeq
The quasi-neutrality condition inside the shock ramp, at equilibrium is written as
\beq
n_{i0}(x)\approx n_{e0}(x).\label{neutrality}
\eeq
Since the equilibrium quasi-neutrality condition just ahead of the shock ramp can be written as $n_{iu}\approx n_{eu}$, then from Eq. (\ref{neutrality}) we can conclude that $\sigma_{in}\approx \sigma_{en}$.

The linear plasma waves can be studied by using the Fourier analysis technique, i.e., by assuming the fluctuating quantities to be of the form $D_1=\tilde{D_1}\exp[i(k_x x+k_z z-\omega t)]$, where $k_x(k_z)$ are the wave vectors along the $x(z)-$directions, and $\omega$ is the angular frequency. Thus, following the linearization scheme discussed above and using the Eq. (\ref{electronmomentPhi}) only at the first order, the Eqs. (\ref{ioncontinuity})-(\ref{ionmomentz}), and (\ref{electronmomentPhi})-(\ref{poisson}) respectively becomes

\begin{equation}
\begin{aligned}
\left[\omega-k_x v_{ix0}(x)-k_z v_{iz0}+i \sigma_{iv}\right]n_{i1}+i\sigma_{in}v_{ix1}-n_{i0}(x)k_z v_{iz1}=0,&\\
\\
\\
n_{i0}(x)\left[\omega-k_x v_{ix0}(x)-k_z v_{iz0}+i \sigma_{iv}\right]v_{ix1}-\frac{k_BT_{i0}(x)}{m_i}\left(k_x+i\frac{\sigma_{in}}{n_{i0}(x)} \right)n_{i1}
+\frac{e}{m_i}n_{i0}(x)k_x\phi_1\\+\frac{1}{m_i}(-n_{i0}(x)+i\sigma_{in})k_BT_{i1}
-i\Omega_in_{i0}(x)v_{iy1}=0,&\\
\\
i\Omega_i v_{ix1}+\left[\omega-k_x v_{ix0}(x)-k_z v_{iz0}\right]v_{iy1}=0,&\\
\\
\left[\omega-k_x v_{ix0}(x)-k_z v_{iz0}\right]v_{iz1}-\frac{e}{m_i}k_z\phi_1=0,&\\
\\
k_BT_{e0}n_{e1}-en_{e0}(x)\phi_1+n_{e0}k_BT_{e1}=0,&\\
\\
\left[\omega-k_xv_{ix0}(x)+i(1-\gamma_i)v_{ix0}(x)\frac{\sigma_{in}}{n_{i0}(x)}  \right]k_BT_{i1}
\\+(1-\gamma_i)\frac{k_BT_{i0}(x)}{n_{i0}(x)}\left[\omega-k_xv_{ix0}(x)
-iv_{ix0}(x)\frac{\sigma_{in}}{n_{i0}(x)}  \right]n_{i1}=0,&\\
\\
k_BT_{e1}+(1-\gamma_e)\frac{k_BT_{e0}(x)}{n_{e0}(x)}n_{e1}=0,&\\
\\
4\pi e n_{i1}-4\pi e n_{e1}-(k_x^2+k_z^2)\phi_1=0.&
\end{aligned}\label{eq1}
\end{equation}
To zero-th order, Eq. (\ref{ioncontinuity}) can be recast as follows
\beq
\frac{\partial}{\partial x}\left[n_{i0}(x) v_{ix0}(x)\right]=0,\quad\mathrm{or}\quad \sigma_{iv}=-\frac{v_{ix0}(x)}{n_{i0}(x)}\sigma_{in}\,.\label{zero-th-continuity}
\eeq
This equation gives the relation between the slopes of ion density and velocity.

\subsection{Parallel waves}
In order to analytically derive the dispersion relation of IAWs, we make some simplifying assumption: the wave is assumed to propagate along the magnetic field (i.e., $k_x=0$ and $k_z\neq0$). Thus, the restoring force provided by the parallel electron pressure is balanced by the ion inertia to excite the parallel propagating IAWs within the ramp. In this case, the electric field is approximately parallel to ${\bf B}$: ${\bf E}_1 = \nabla \phi \simeq -i{\bf k} \phi_1 = (0,0,-i k_z \phi_1)$. This electrostatic field adds to the motional electric field (-${\bf v} \times {\bf B}$) in accelerating the charged particles at the shock. However, we do not consider the particle acceleration in this paper.

At 1 au, the ion gyrofrequency $\Omega_i/2 \pi \approx0.0748$ Hz
(using $B_0\approx5\,n$T) is negligibly small as compared to the frequency of IAWs (i.e., $\omega\gg\Omega_i$) being studied which is of the order of kHz \citep{Hess1998,Wilson2007}. It is also found that $\Omega_i/2 \pi<<\omega_{pi}/2 \pi \sim 2$ kHz and $\Omega_i/2 \pi<<\omega_{pe}/2 \pi \sim 80$ kHz. These relations also hold closer to the Sun ($0.25$ au).
Considering the plasma to be in equilibrium before the shock arrival leads us to $(v_{iy0},v_{iz0},\phi_0)\approx0$.
After a lengthy algebraic manipulation of Eqs. (\ref{eq1}), a fourth-degree polynomial is obtained, with four unique solutions (see Appendix \ref{Appendix-A}):

\beq
\omega^4+iP_3\omega^3+P_2(x)\omega^2+iP_1(x)\omega+P_0(x)=0,\label{dr}
\eeq
where
\beq
P_0(x)=- \sigma_{iv}^2\gamma_e(1-\gamma_i)\frac{ C_{si}^2(x)k_z^2}{(1+\gamma_ek_z^2\lambda_D^2(x))},
\nonumber
\eeq
\beq
P_1(x)= \sigma_{iv}\left[(1-\gamma_i) \sigma_{iv}^2-\gamma_i\gamma_e\frac{ C_{si}^2(x)k_z^2}{(1+\gamma_ek_z^2\lambda_D^2(x))}\right],
\nonumber
\eeq
\beq
P_2(x)=-\left[(2\gamma_i-1) \sigma_{iv}^2+(2-\gamma_i)\frac{ \sigma_{in}^2}{{n_{i0}^2(x)}}\frac{k_BT_{i0}(x)}{m_i}  +\frac{\gamma_e C_{si}^2(x)k_z^2}{(1+\gamma_ek_z^2\lambda_D^2(x))}\right],
\nonumber
\eeq
\beq
P_3= \sigma_{iv} (1+\gamma_i),
\nonumber
\eeq
\beq
C_{si}^2(x)=\lambda_D^2(x)\,\omega_{pix}^2(x)=\frac{k_BT_{e0}(x)}{m_i},
\label{Csi}
\eeq
and
\beq
\lambda_D^2(x)=\frac{k_BT_{e0}(x)}{4\pi n_{i0}(x)e^2}, \,\, \omega_{pix}^2(x)=\frac{4\pi n_{i0}(x) e^2}{m_i}.\nonumber
\eeq

By looking at the solutions $\omega_{1-4}$, given in Eqs. (\ref{exact-sol-1,2}-\ref{exact-sol-3,4}), it is hard to recognize the wave modes. So, in order to simplify these solutions, we have to make some assumptions. Ions are assumed to be ideal gas with specific heat ratio $\gamma_i=\frac{5}{3}$. For electrons within the ramp, while considering the high frequency of IAWs ($\geq$ kHz), we can assume $\gamma_e=1$ (isothermal)
due to their large thermal speed:
for $T_{eu}=12$ eV (at $\approx1$ au), the electron thermal speed is 3403.97 $\rm km\,s^{-1}$, and for $T_{eu}=34.4$ eV (at $\approx0.25$ au), electron thermal speed is 4589.9 $\rm km\,s^{-1}$, both far larger than the phase velocity of IAWs \citep{Nicholson1983}
($\sim 50$ $\rm km\,s^{-1}$, see Figs. \ref{Vph1-rin} and \ref{Vph1-ret}). Also, defining
\beq
A(x)=\frac{C_{si}(x)^2k_z^2}{(1+k_z^2\lambda_D(x)^2)}\quad \quad \mathrm{and}\quad \quad L(x)=\frac{ \sigma_{in}^2}{{n_{i0}(x)^2}} \frac{k_BT_{i0}(x)}{m_i},
\eeq
it is found that $A(x)>>\sigma_{iv}>>L(x)$, throughout the parameter range considered herein. By keeping the two highest dominant terms, the simplified forms of $\digamma$ and $\ell$ (full forms given in Eqs. (\ref{full-F}-\ref{full-Ell})) can be written respectively as:

\beq
\digamma(x)=-A(x)\left(1-{\frac{\sqrt{3}}{A(x)^{1/2}}}\,|(\gamma_i-2)\sigma_{iv}|  \right)\quad\quad \mathrm{and} \quad \quad\ell=i\,\,|(\gamma_i-2)\sigma_{iv}|.
\eeq
Using these simplified expressions of $\digamma$ and $\ell$ into $\omega_{1-4}$ (Eqs. (\ref{exact-sol-1,2}-\ref{exact-sol-3,4})), and again keeping only the dominant  terms and neglecting the smaller terms, we get the simplified forms of $\omega_{1-4}$ respectively:

\begin{eqnarray}
\omega_{1,2}(x)=\pm\frac{1}{2}\left[4\frac{C_{si}(x)^2k_z^2}{(1+k_z^2\lambda_D(x)^2)}-\frac{{|\sigma_{iv}|}}{3\sqrt{3}} \frac{C_{si}(x)k_z}{(1+k_z^2\lambda_D(x)^2)^{1/2}}
 \right]^{1/2}+\,i\,\frac{|\sigma_{iv}|}{2},\label{Ap-sol-1,2}
\label{eq:omega1}
\end{eqnarray}

\begin{multline}
\omega_{3,4}(x)=i\,\frac{5}{6}\,|\sigma_{iv}|\pm i\frac{1}{2}\Biggl[-\frac{C_{si}(x)^2k_z^2}{(1+k_z^2\lambda_D(x)^2)}
\Biggl\{-\frac{1}{3}+\Biggl(\frac{3\,C_{si}(x)^2k_z^2}{(1+k_z^2\lambda_D(x)^2)}
+2\frac{ \sigma_{in}^2}{{n_{i0}(x)^2}}  \frac{k_BT_{i0}(x)}{m_i}-2\sigma_{iv}^2\Biggr)\\ \times \left(\frac{9\,C_{si}(x)^2k_z^2}{(1+k_z^2\lambda_D(x)^2)}
-\frac{3\sqrt{3}\,\,C_{si}(x)k_z}{(1+k_z^2\lambda_D(x)^2)^{1/2}}
\,\,|\sigma_{iv}|\right)^{-1}\Biggr\}
+\frac{C_{si}(x)k_z}{(1+k_z^2\lambda_D(x)^2)^{1/2}}
\,\,\frac{|\sigma_{iv}|}{3\sqrt{3}}
 \Biggr]^{1/2}.\label{Ap-sol-3,4}
\end{multline}

The two frequencies $\omega_{1,2}$ are complex with real part of opposite sign whereas the two frequencies $\omega_{3,4}$ are purely imaginary modes.
Equations (\ref{Ap-sol-1,2})-(\ref{Ap-sol-3,4}) show that all the four solutions $\omega_{1-4}$ have a finite growth rate due to the jump in ion velocity across the shock. In other words, we find that the IAWs instability is directly generated by the shock transition.

The dispersion relation in Eqs. (\ref{Ap-sol-1,2})-(\ref{Ap-sol-3,4}) presents significant differences from the textbook dispersion relation \citep[e.g.,][]{Nicholson1983} for parallel propagating IAWs (with $\omega>>\Omega_{i}$) in a homogeneous medium
\beq
\omega^2=k_z^2\gamma_i \frac{k_BT_i}{m_i}+\frac{\gamma_e k_z^2k_BT_e/m_i}{1+\gamma_ek_z^2\lambda_D^2},
\eeq
where $\lambda_D$ is the Debye length and $m_i$ is the ion mass. The kinetic-scale condition for IAW instability in hot plasmas remarkably depends on the electron-to-ion temperature ratio and on $k\lambda_D$, in the two cases that the wave growth is generated either via parallel electric current \citep[Eq. 5.2.40 therein]{Benz2002} or by ion beams \citep{Treumann1997}; in both the cases, the instability is triggered provided that the $T_e/T_i > 1$, with a saturation at $T_e \simeq 10\, T_i$, and that $(k\lambda_D)^2 \ll 1$. However, such a $T_e/T_i$-dependence has been found to be inconsistent with a number of high-time resolution in-situ measurements
\citep[e.g.,][]{Goodrich2019,Vech2021}.

The fluid-approach presented herein shows that a shock compression modifies the instability threshold so that even near the electron-ion equilibration, rarely realized at IP and astrophysical shocks, IAWs can grow, as suggested by recent MMS in-situ measurements (e.g., \citet{Goodrich2018}).

The current- or beam-driven parallel instabilities are not direct analogs of our parallel instability that is triggered by the plasma compression, and hence the high-frequency instability across the shock transition found here can be considered a new type of instability.

A model of propagation of IAWs in inhomogeneous media in linear regime \citep{Doucet1974} found, for isothermal compressions, that the wave amplitude scales with the square root of the unperturbed density, so it increases within compression regions. We find that the growth rate depends on the shock compression $r_{in}$ and shock speed, whereas the wave frequency depends on $\sigma_{in}$ only if the condition $C_{si} k_z \gg \sigma_{iv}$ does not hold as the second term in the real part of Eq. (\ref{Ap-sol-1,2}) will be not negligible. In the following section we investigate the effect of different plasma and IP shock parameters on the properties of the IAWs.

\section{Ion acoustic waves instability at interplanetary shocks\label{Instability}}
The analysis of the solutions given in Eqs. (\ref{Ap-sol-1,2})-(\ref{Ap-sol-3,4}) shows that the wave frequency is complex and has the form $\omega=\omega_{r}+i\gamma_{}$, where $\omega_{r}$ is the real part of the frequency and $\gamma_{}$ is the imaginary part. If $\gamma_{}$ is positive, the wave grows and becomes unstable, and $\gamma_{}$ is called the growth rate of instability. On the other hand, if $\gamma_{}$ is negative, the wave damps over time until the amplitude dies out.

Now, considering the wave frequency corresponding to the solution $\omega_1$,  Eq. (\ref{Ap-sol-1,2}), with the real part $\omega_{1r}$, and imaginary part $\gamma_{1}$, we can write (after dropping the explicit dependence on `$x$')
\beq
\omega_1=\omega_{1r}+i\gamma_{1},\label{omega1-r-i}
\eeq
where
\begin{eqnarray}
\omega_{1r}=\frac{1}{2}\left[\frac{4\,\,C_{si}^2k_z^2}{(1+k_z^2\lambda_D^2)}-\frac{{|\sigma_{iv}|}}{3\sqrt{3}} \frac{C_{si}k_z}{(1+k_z^2\lambda_D^2)^{1/2}}  \,\,
 \right]^{1/2} \quad\mathrm{and}
 \quad \gamma_1 = \frac{v^{sh}_{ixu}}{2}\left|\frac{1}{r_{in}}-1 \right|\frac{\omega_{pi}}{c}.
 \label{eq:omega1_gamma1}
\end{eqnarray}
It should be noticed that the growth rate $\gamma_1$ of the IAWs is independent of the wavenumber $k_z$.
The properties of IAWs corresponding to the dispersion relation (\ref{omega1-r-i}) can be measured in-situ. We notice that with the particular choice of parallel wave propagation direction made herein, the Doppler-shifted measured wave frequency $\omega_{1,sc}$ is approximately equal to the frequency in the solar wind frame $\omega_1$ \citep{Hess1998}:
\beq
\omega_{1,sc}=\omega_1+{\bf k}.{\bf V_{sw}},\label{doppler1}
\eeq
where ${\bf k} = (0,0,k_z)$ is the wave vector, ${\bf V_{sw}} \simeq (V_{sw}, 0,0)$ is the approximate solar wind speed, and the Taylor hypothesis for electric field fluctuations frozen into the plasma is used.
Since ${\bf k}=k_z \hat{z}$ and ${\bf V_{sw}}\simeq V_{sw}\,\hat{x}$, the second term in (\ref{doppler1}) vanishes and the spacecraft observes the frequency $\omega_{1,sc}=\omega_1$. The measured Doppler shift term is comparable to the rest frame frequency in the quiescent solar wind, i.e., of the order of kHz \citep{Gurnett1979b,Gurnett1979,Pisa.etal:21}.

Previous studies \citep[e.g.,][]{Gurnett1979b,Gurnett1979,Wilson2007} reported IAW observation with frequencies
of the order of $\sim$ few kHz at shocks. For solar wind speed at $1$ au in the range $V_{sw}=400-800$ $\rm km\,s^{-1}$,
the wavenumber $k_z \sim \omega_{1,sc}/V_{sw}$
is expected to be in the range  $10^{-5}-10^{-3}\,$ cm$^{-1}$, much larger than $(c/\omega_{pi})^{-1} \simeq 10^{-7}\,$ cm$^{-1}$ and therefore with a length scale much smaller than the shock ramp thickness consistent with the impulsive nature of IAWs as it appears in high-time resolution MMS in-situ measurements \citep{Goodrich2018}.

Equations (\ref{Ap-sol-1,2}) and (\ref{Ap-sol-3,4}) show that both real and imaginary parts of the IAW frequency are functions of the shock strength responsible for the wave excitation. Figure \ref{omega1-rin} shows the $\omega_{1r}$ vs $k_z$ relation for different values of $r_{in}$. In Fig. \ref{omega1-rin}, at the spatial location $x=0.5\,{c}/{\omega_{pi}}$ within the ramp, we used: $r_{in}=2.0$, $r_{it}=10$, $r_{et}=5$, $n_{iu}=5$ $\mathrm{cm^{-3}}$, $v_{ixu}=4\times10^7 \mathrm{\,cm \,s^{-1}}$, $k_BT_{iu}=11\,{\rm eV}$ and $k_BT_{eu}=12\,{\rm eV}$ for the blue solid curve. For the red (black) solid curves we used $r_{in}=2.5$ ($r_{in}=3.5$) leaving unchanged the other parameters. It can be observed that a $\sim\,$kHz wave frequency increases with the increase in shock strength $r_{in}$. In Fig.  \ref{omega1-rin}, we also plotted the ion plasma frequency $\omega_{pix}$ (dashed curves) within the shock ramp, and that in the upstream region $\omega_{pi}$ (green-dotted curves).
It is observed that the ion plasma frequency within the shock ramp increases with the increase in shock strength $r_{in}$. A comparison between the green-dotted curve (for $\omega_{pi}$) and the three dashed curves (for $\omega_{pix}$) reveals that the ion plasma frequency is higher inside the shock ramp as compared to that in the upstream region, as expected: higher ion plasma frequency results from the increased plasma number density inside the shock ramp caused by the shock compression.
In Fig. \ref{Vph1-rin}, we plotted the phase velocity $V_{ph1}={\omega_{1r}}/{k_z}$ for different values of the shock strength $r_{in}$. It is observed that the phase velocity of the IAWs also increases with the increase in shock strength. This behavior can be due to the fact that the IP shocks act as a free energy source, thereby affecting the dispersion relation of the IAWs \citep{Treumann1997}, and as the strength of the IP shock is increased, more free energy becomes available to generate faster IAWs.

\begin{figure}[h]
\centering
\subfigure[$\omega_{1r}$ (solid curves), $\omega_{pix}$ (dashed curves) and $\omega_{pi}$(green-dotted curve)]{\label{omega1-rin}
\includegraphics[width=0.45\linewidth]{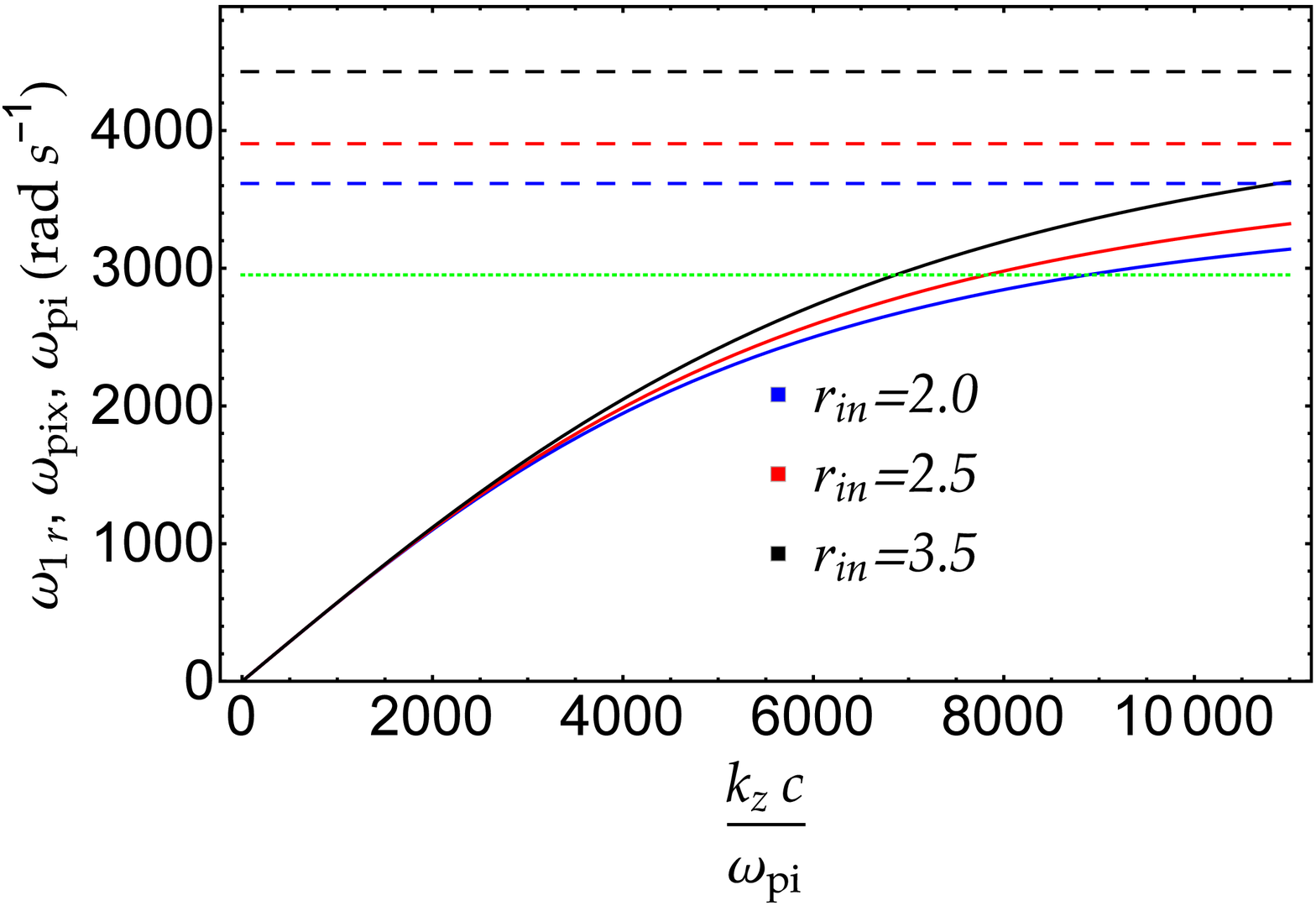}}
\subfigure[Phase velocity $V_{ph1}$]{\label{Vph1-rin}
\includegraphics[width=0.45\linewidth]{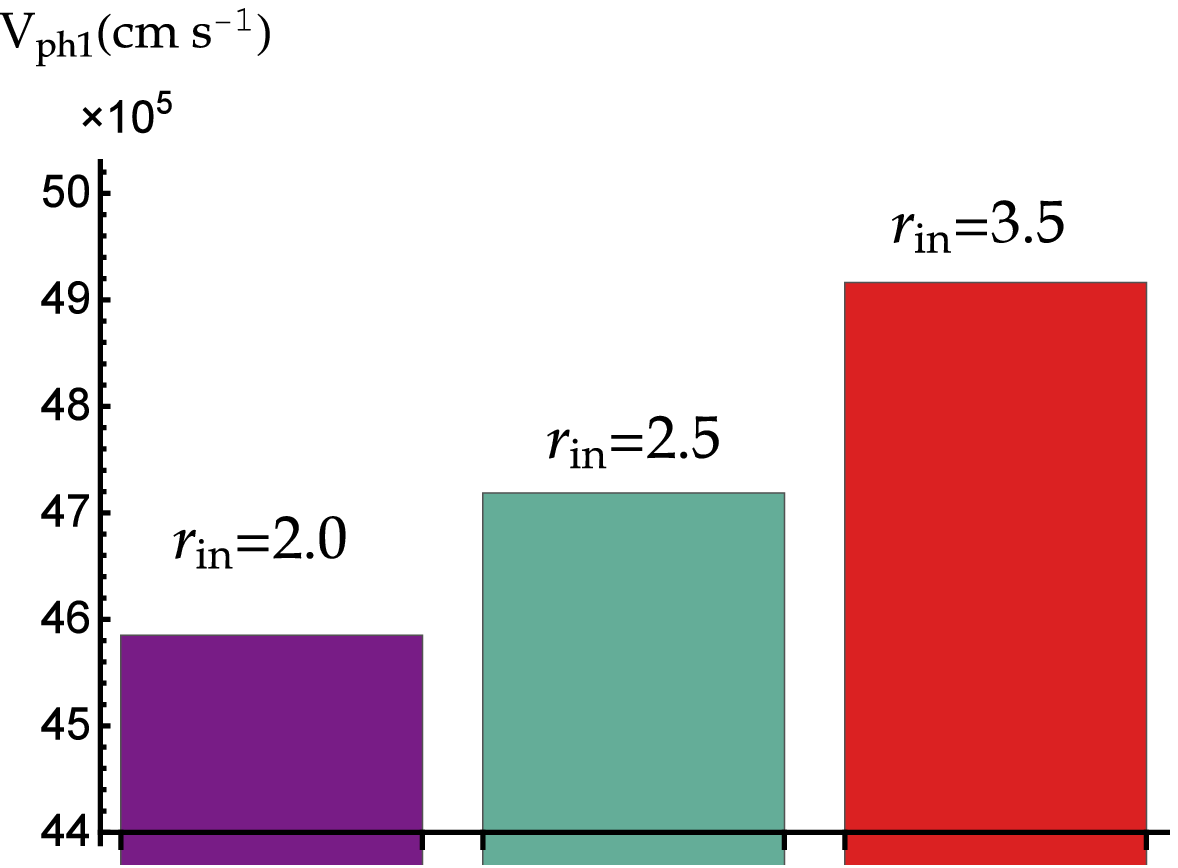}}
\qquad
\subfigure[Group Velocity $V_{gp1}$]{
\label{group-rin}
\includegraphics[width=.5\linewidth]{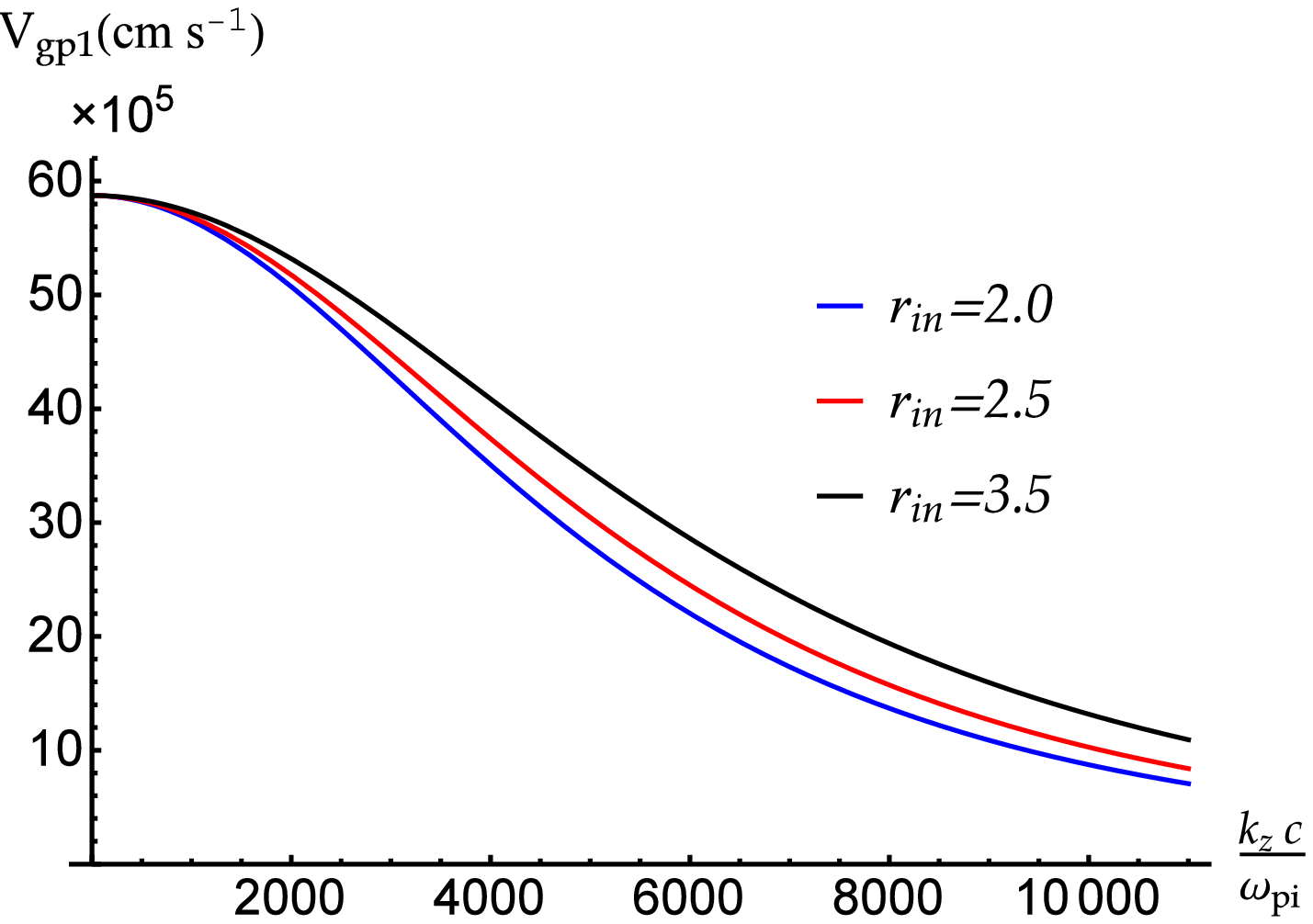}}
\subfigure[$\gamma_1$ vs $r_{in}$]{\label{sigma1-rin}
\includegraphics[width=.45\linewidth]{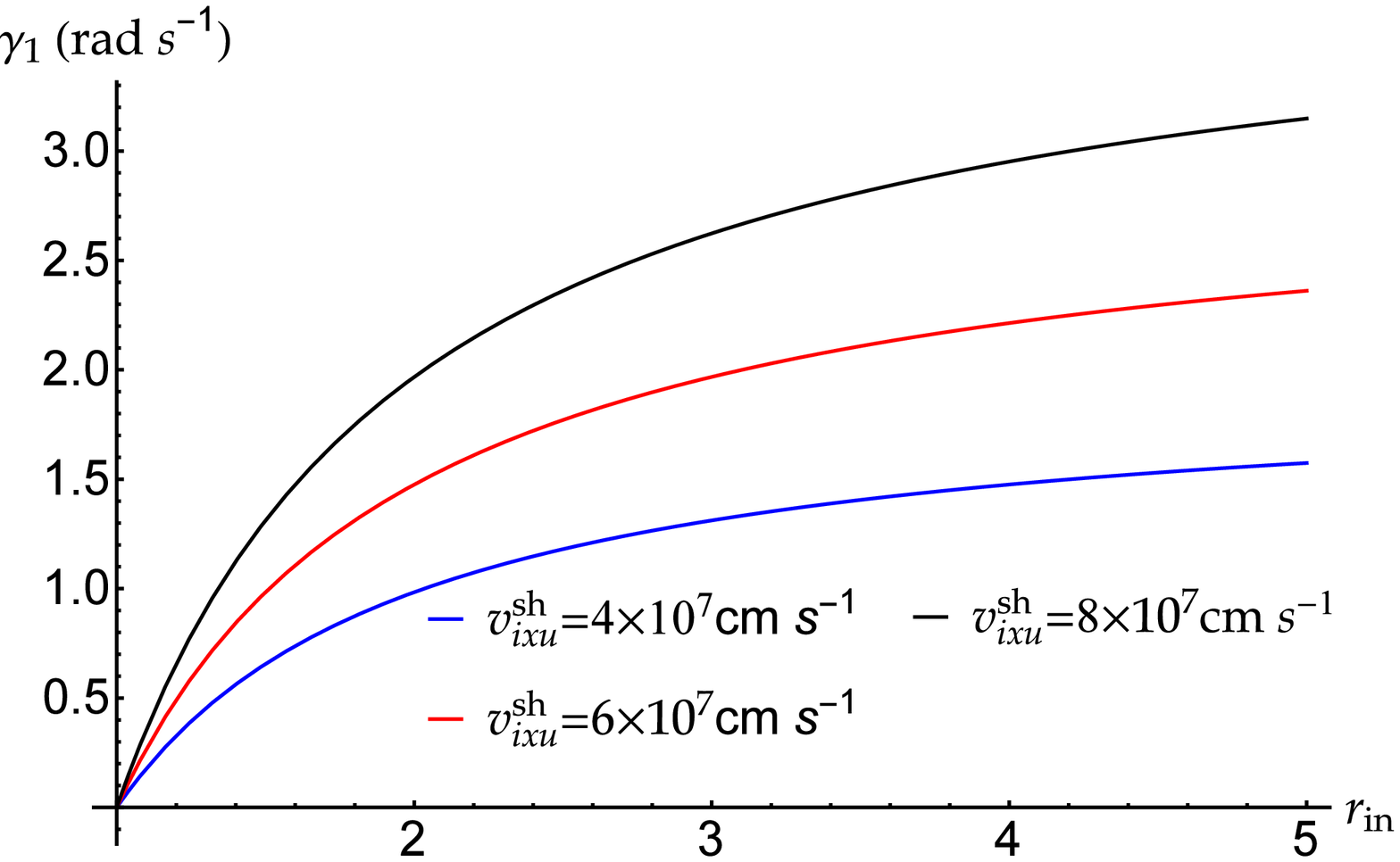}}
\centering
\caption{The effect of increase in shock strength $r_{in}$ at the location $x=0.5\,{c}/{\omega_{pi}}$ within the ramp. For the blue curves in panels (a), (c) and (d): $r_{in}=2$, $r_{it}=10$, $r_{et}=5$, $n_{iu}=5$ cm$^{-3}$, {$v^{sh}_{ixu}=4\times10^7$ cm\,s$^{-1}$} and $k_BT_{iu}=11\,\rm eV$, $k_BT_{eu}=12\,\rm eV$. For panel (b), $\frac{k_z c}{\omega_{pi}}=5000$, and the remaining parameters are the same as used for blue curve in panel (a).} \label{rin}
\end{figure}

The group velocity
\beq
V_{gp1}=\frac{\partial \omega_{1r}}{\partial k_z}
\eeq
of IAWs is influenced by the increase in shock compression $r_{in}$ as shown in Fig. \ref{group-rin}. At a fixed value of the wave number $k_z$, the group velocity increases with the increase in shock strength $r_{in}$. Moreover, $V_{gp1}$ is smaller than the IP shock speed; thus, the IAWs generated within the shock ramp are quickly convected downstream and cause negligible corrugation of the shock surface and are unlikely to contribute significantly to the upstream self-generated turbulence although they might strongly scatter charged particles within the ramp.

In Fig. \ref{sigma1-rin}, we plotted the imaginary part of the dispersion relation $\gamma_1$ vs $r_{in}$ for different values of the shock speeds.  Equation (\ref{eq:omega1}) shows that at sufficiently long wavelengths, i.e., at $k_z\approx0$, the wave turns into a purely growing mode with no oscillatory behavior. It is important to mention that the IAW instability studied so far satisfies the linear instability condition \cite[sect. 2.1,][]{Treumann1997}, according to which the growth rate $\gamma_1$ should be sufficiently smaller than the real frequency $\omega_{1r}$, i.e.,
\beq
\frac{\gamma_1}{\omega_{1r}}\ll1 \, ,
\eeq
hence justifying our linear approximation.
It is also evident from the expression of the growth rate, $\gamma_1={|\sigma_{iv}|}/{2}$, that the instability is excited solely by the free energy provided by the shock itself. For more insight, we can see that if there is no shock, i.e., for $r_{in}\rightarrow1$, $|\sigma_{iv}|\rightarrow0$, and $\gamma_1\rightarrow0$. It is observed that the wave growth rate $\gamma_1$ increases with the increase in shock compression ratio $r_{in}$. This behavior is in agreement with the Wind measurements \citep[Figure 2 therein]{Wilson2007} that show a positive correlation between the IAW wave amplitude and the shock compression ratio $r_{in}$. The wave amplitude as high as 300 $\rm mV m^{-1}$ were seen at $r_{in}\sim3$.

It is also important to mention here that the IAW's growth rate $\gamma_1$ is unaffected by the electron-to-ion temperature ratio for the upstream region ($T_{eu}/T_{iu}$), or within the shock ramp ($T_{e0}(x)/T_{i0}(x)$).
This result is in agreement with the observations of the IAWs associated with the IP shocks \citep{Hess1998,Wilson2007} and Earth's bow shock crossings \citep{Goodrich2016,Goodrich2018}, where the IAWs with growing features are observed although the condition
$T_{eu}/T_{iu}>>3$
for instability is never satisfied. Equation (\ref{eq:omega1_gamma1}) also shows that the increase in shock speed $v_{ixu}^{sh}$ leads to an increase in the growth rate that likely leads to the generation of higher amplitude and higher frequency IAWs because the faster shocks supply more free energy. A further parametric analysis also revealed that the jump in ion temperature $r_{it}$ within the shock ramp has no effect on the frequency ($\omega_{1r}$), the phase velocity ($V_{ph1}$) or the group velocity ($V_{gp1}$).
Since the wave frequency $\omega_{1r}$, as well as $\gamma_1$, also depends on the $|\sigma_{iv}|$, as a consistency check we verified that in the absence of compression ($r_{in} = 1$) the dispersion relation of the standard IAWs is retrieved \citep{Baumjohann1997}:
\beq
\omega_{1r}=A=\frac{C_{si}^2k_z^2}{(1+k_z^2\lambda_D^2)}.
\eeq

\begin{figure}[h]
\centering
\subfigure[$\omega_{1r}$ vs $k_z$]{\label{omega1-ret}
\includegraphics[width=.485\linewidth]{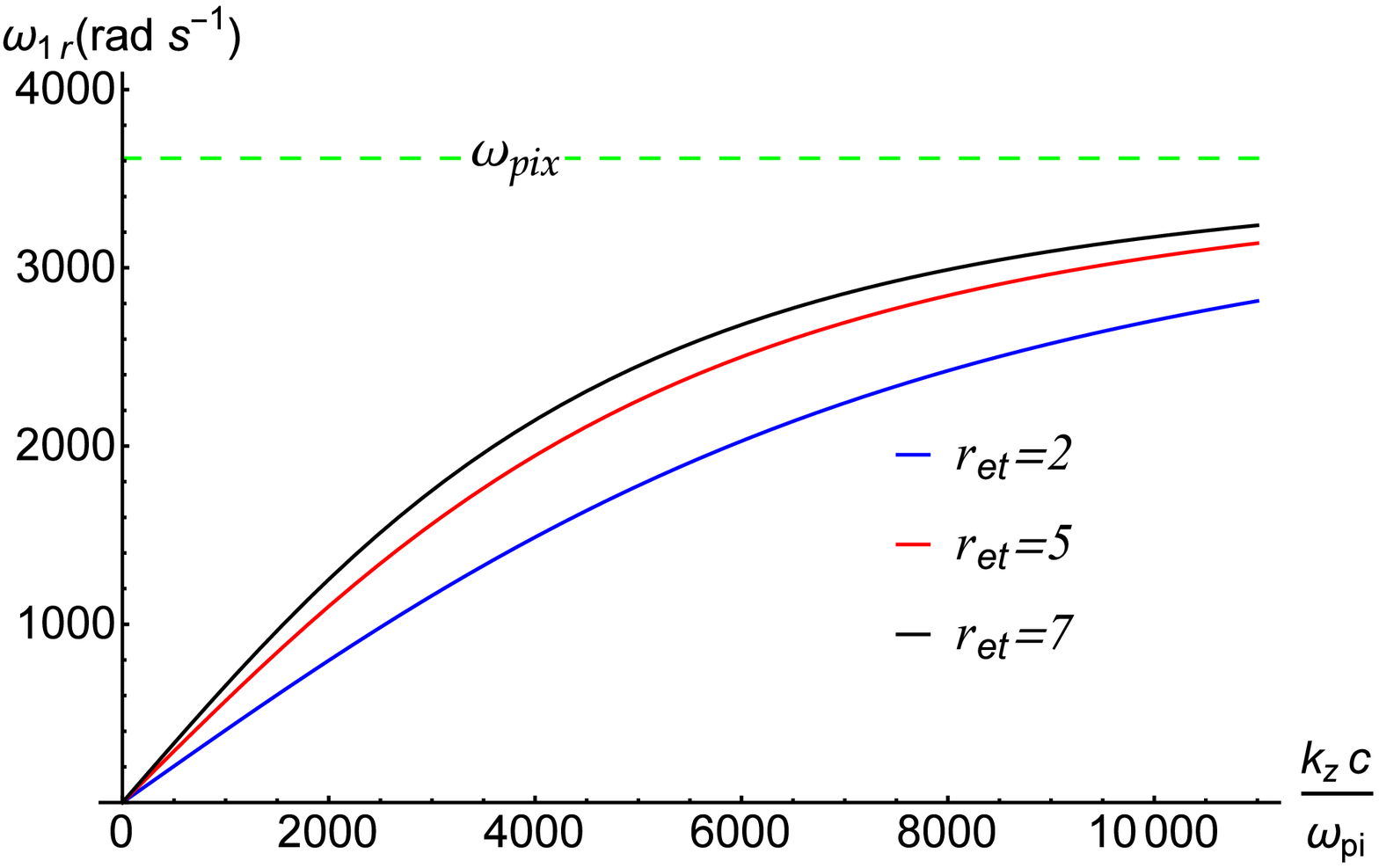}}
\subfigure[The phase velocity $V_{ph1}$]{\label{Vph1-ret}
\includegraphics[width=.485\linewidth]{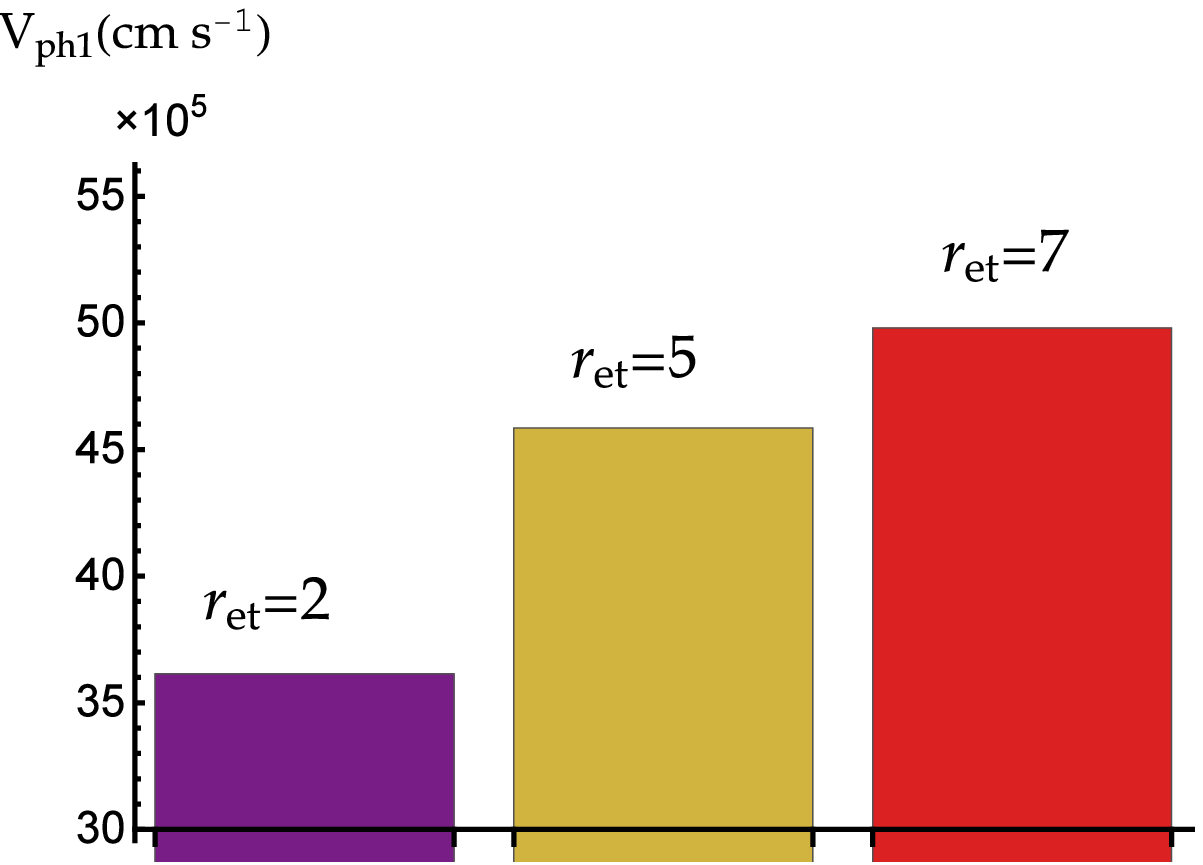}}

\subfigure[The group velocity $V_{gp1}$]{\label{group-ret}
\includegraphics[width=.485\linewidth]{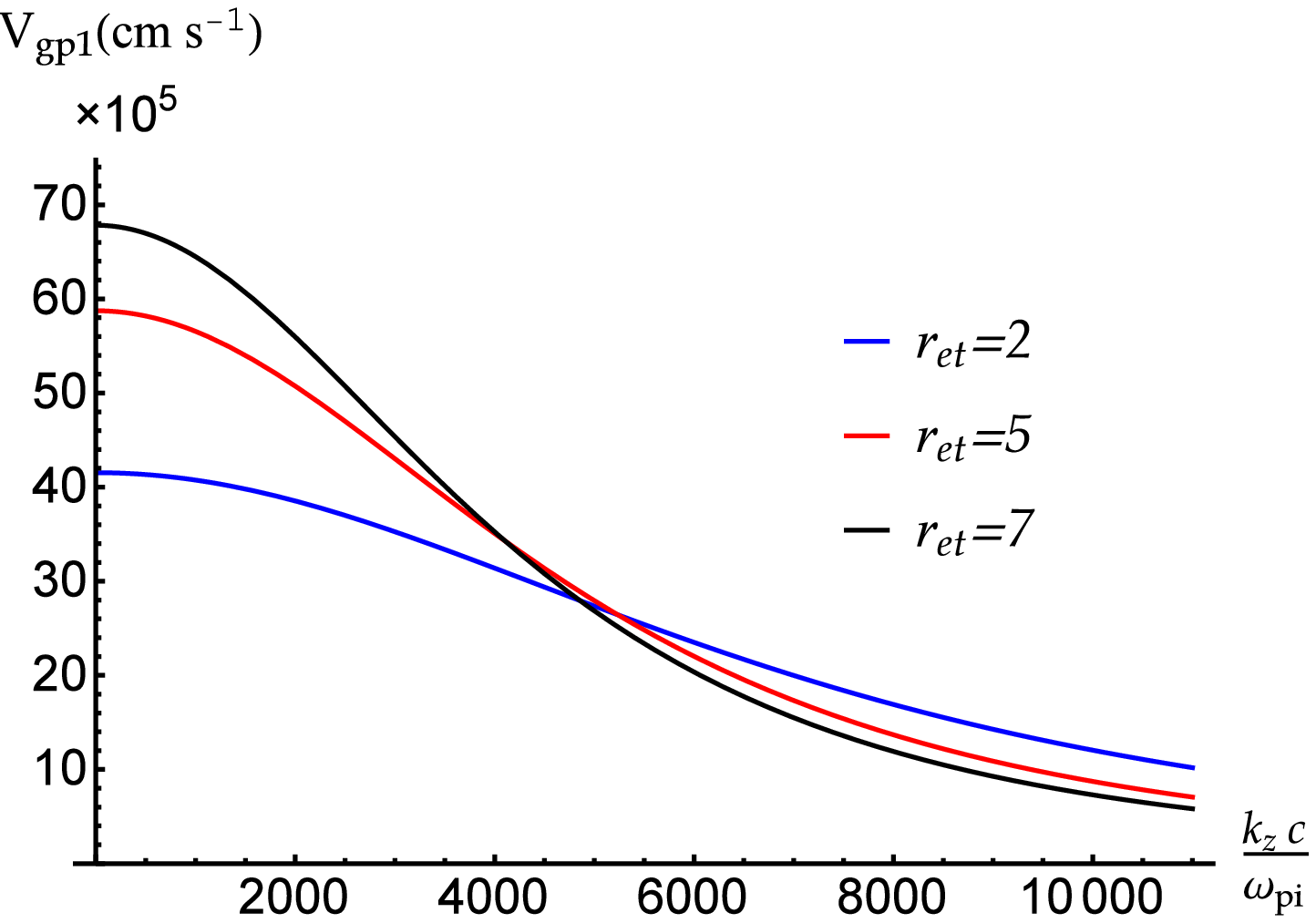}}
\centering
\caption{The effect of increase in electron temperature jump $r_{et}$ at $x=0.5\frac{c}{\omega_{pi}}$, with remaining parameters same as given in Fig. \ref{rin}.} \label{ret}
\end{figure}
In Fig. \ref{ret}, the effect of increase in electron temperature jump $r_{et}$ on the frequency, the phase velocity, and the group velocity of the IAWs are plotted, for fixed values of the remaining parameters.
Figure \ref{omega1-ret} shows that at larger length scales (i.e., small $k_z$ or $k_z \, c/\omega_{pi} \lesssim 10^4 $), the frequency $\omega_{1r}$ increases with the increase in $r_{et}$ for a given $k_z$, but at shorter scales (i.e., large $k_z$ or  $k_z \, c/\omega_{pi} \gtrsim 2 \times 10^4$) the effect of increase in $r_{et}$ becomes less pronounced. An increase in electron heating (via $r_{et}$) at the shock also has a significant effect on the phase velocity $V_{ph1}$ of the IAWs as it can be seen from the Fig. \ref{Vph1-ret}. The increase in $r_{et}$ leads to higher phase velocity of the IAWs.

The group velocity shows an interesting behavior when we increase the electron temperature jump $r_{et}$, as shown in Fig. \ref{group-ret}. At larger length scales (i.e., at small $k_z$ or $k_z \, c/\omega_{pi} \lesssim 2 \times 10^3 $) the group velocity is relatively large and increases with the increase in $r_{et}$; at small scales (i.e., at large $k_z$ or $k_z \, c/\omega_{pi} \gtrsim 8 \times 10^3 $) the group velocity slightly decreases or nearly becomes independent of the jump in electron temperature. This trend may be owing to the fact that the increase in electron heating offers more free energy for increased effective speed $C_{si}(x)$
within the shock ramp, which then leads to higher phase and group velocities.
Since the ion plasma waves are independent of the electron heating, so the increase in $r_{et}$ at larger $k_z$ has a very small effect on the frequency, the phase velocity, and the group velocity of IAWs. The growth rate $\gamma_1$ is independent of the temperature jump $r_{et}$, hence not plotted.

\begin{figure}[h]
\centering
\includegraphics[width=.5\linewidth]{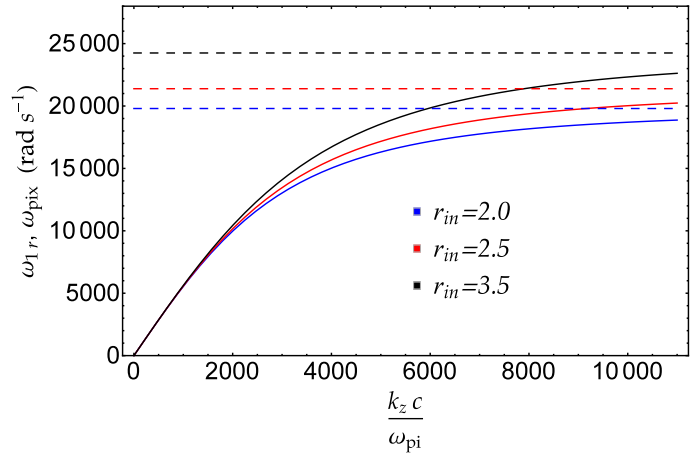}
\centering
\caption{The ion plasma frequency $\omega_{pix}$ (dashed curves) and $\omega_{1r}$ (solid curves), for different $r_{in}$ at $0.25$ au, with
$k_BT_{iu}\approx34.46{\,\rm eV}$, $k_BT_{eu}\approx40$ eV, and $n_{iu}=150\, \mathrm{cm^{-3}}$.
The remaining parameters are same as in the caption of Fig. \ref{rin} for blue curve. } \label{0.25AU1}
\end{figure}

We have also applied the model to measurements  closer to the Sun. The high-frequency ($\sim$kHz) electric field measurements by SolO at 0.8 au of the 2021 November 3 shock event, for instance, show excitation of electric field modes with laminar magnetic field in the same frequency range, that suggest electrostatic modes in the IAWs expected frequency range (Thomas Chust and Jan Soucek, Private communication).
It is expected that the IAW activity may be enhanced near the Sun due to the enhanced temperatures, densities, and turbulence therein (also larger $\omega_{pix}$). In order to find the behavior of IAWs at 0.25 au, where $k_BT_{iu}\approx34.46$ eV, $n_{iu}=150\, \mathrm{cm^{-3}}$ \citep{Bale2019,Kasper2019} and $k_BT_{eu}\approx40$ eV \citep{Moncuquet2020}, we observe (see Fig. \ref{0.25AU1}) that the IAWs have significantly high frequency $\omega_{1r}$ as compared to the IAWs near 1 au, due to the temperature scaling through the solar wind (see Eq. (\ref{Csi})).}

\section{Conclusions\label{conclusion}}
In this paper, we have built a fluid-scale model to investigate the generation of IAWs within the ramp regions of the perpendicular IP shocks by using a novel linearization method. In a two-fluid approach (electrons/ions), we linearize the fluid equations coupled with the Poisson equation for the electrostatic potential inside the shock ramp. For parallel propagating waves, a linear wave decomposition yields four complex solutions which corresponds to four different electrostatic IAW modes. By performing a parametric analysis within the in-situ measurement ranges of the solar wind, we find two roots purely imaginary that represents purely growing modes. The other two roots are complex with positive imaginary parts,  hence finite growth rate. The frequency falls in the $\sim$kHz to $\sim$tens of kHz range in the spacecraft frame, consistent with, e.g., MMS and SolO measurements in multiple solar wind locations (at $1$ au and also at $0.25$ au). In contrast to previous kinetic scale models and in-situ measurements, the wave growth rate is not affected by the electron-to-ion temperature ratio, nor by ion heating across the shock transition, $r_{it}$; additionally, the growth rate increases with shock density compression, as suggested by Wind measurements, and with shock speed. 

Disentangling wave modes within the shock self-generated turbulence that scatters energetic particles very close to the shock has been intensively attempted by the small scale particle-in-cell simulations in the past decade \citep[for a review see][]{Pohl.etal:20}. One of our findings is that the group velocity of the parallel electrostatic modes is much smaller than the shock speed; as a result, at perpendicular shocks a shock corrugation by these electrostatic modes is expected over a relatively small patch of the shock surface.

The model presented here provides a framework for the generation of high-frequency electrostatic waves within a variety of inhomogeneous plasmas, not only shock compressions, and might be more broadly applied to the ubiquitous ion-acoustic fluctuations detected by SolO in the inherently inhomogeneous quiescent solar wind \citep{Pisa.etal:21}.

\begin{acknowledgments}
We thank the referee for useful and constructive comments. This work was supported, in part, by NASA under Grants NNX15AJ71G and 80NSSC18K1213, and by NSF under grant 1850774. FF was supported, in part, also by NASA through Chandra Theory Award Number $TM0-21001X$, $TM6-17001A$ issued by the Chandra X-ray Observatory Center, which is operated by the Smithsonian Astrophysical Observatory for and on behalf of NASA under contract NAS8-03060.
\end{acknowledgments}

\appendix
\section{Solutions of Polynomial in Equation (28)} \label{Appendix-A}
Solving the fourth degree Eq. (\ref{dr}), the linear dispersion relations of the IAWs in the IP shock ramp is obtained. After dropping the explicit dependence on `$x$', these solutions can be written as:
\beq
\omega_{1,2}=-i \frac{P_3}{4}-\frac{1}{2}\ell
\pm\frac{1}{2}\biggl(-\frac{4P_2}{3}-\frac{P_3^2}{2}-\frac{(12P_0+P_2^2+3P_1P_3)}{3\digamma}
-\frac{\digamma}{3}
-\frac{i(-8P_1+4P_2P_3+P_3^3)}{4\ell}\biggr)^{1/2},\label{exact-sol-1,2}
\eeq

\beq
\omega_{3,4}=-i \frac{P_3}{4}+\frac{1}{2}\ell
\pm\frac{1}{2}\biggl(-\frac{4P_2}{3}-\frac{P_3^2}{2}-\frac{(12P_0+P_2^2+3P_1P_3)}{3\digamma}
-\frac{\digamma}{3}
+\frac{i(-8P_1+4P_2P_3+P_3^3)}{4\ell}\biggr)^{1/2},\label{exact-sol-3,4}
\eeq
where
\begin{multline}
\digamma=\frac{1}{2^{1/3}}\Biggl(-27 P_1^2-72P_0 P_2+2P_2^3+9P_1P_2P_3-27P_0P_3^2\\
+\biggl\{-4(12P_0+P_2^2+3P_1P_3)^3+(-27P_1^2-72P_0P_2+2P_2^3+9P_1P_2P_3-27P_0P_3^2)^2\biggr\}^{1/2}\Biggr)^{1/3}\label{full-F}
\end{multline}
and
\begin{equation}
\ell=\biggl(-\frac{2P_2}{3}-\frac{P_3^2}{4}+\frac{(12P_0+P_2^2+3P_1P_3)}{3\digamma}+\frac{\digamma}{3}\biggr)^{1/2}.\label{full-Ell}
\end{equation}

\bibliography{Final-ref}{}
\bibliographystyle{aasjournal}

\end{document}